# Quantum Machine Learning for Optimizing Entanglement Distribution in Quantum Sensor Circuits

Laxmisha Ashok Attisara
*Department of Computer Science*
*Cleveland State University*
Cleveland OH, 44115, U.S.A.
l.ashokattisara@vikes.csuohio.edu

Sathish A. P. Kumar
*Department of Computer Science*
*Cleveland State University*
Cleveland OH, 44115, U.S.A.
s.kumar13@csuohio.edu

*Abstract* -- **In the rapidly evolving field of quantum computing, optimizing quantum circuits for specific tasks is crucial for enhancing performance and efficiency. More recently, quantum sensing has become a distinct and rapidly growing branch of research within the area of quantum science and technology. The field is expected to provide new opportunities – especially regarding high sensitivity and precision. Entanglement is one of the key factors in achieving high sensitivity and measurement precision [3]. This paper presents a novel approach utilizing quantum machine learning techniques to optimize entanglement distribution in quantum sensor circuits. By leveraging reinforcement learning within a quantum environment, we aim to optimize the entanglement Layout to maximize Quantum Fisher Information (QFI) and entanglement entropy, which are key indicators of a quantum system's sensitivity and coherence with minimization of circuit depth and gate counts. Our implementation, based on Qiskit, integrates noise models and error mitigation strategies to simulate realistic quantum environments. The results demonstrate significant improvements in circuit performance and sensitivity, highlighting the potential of machine learning in quantum circuit optimization by measuring high QFI and entropy in the range of 0.84-1.0 with depth and gates count reduction by 20- 86%.**

## I. Introduction And Background

Quantum computing has emerged as a transformative technology with the potential to solve complex problems beyond the reach of classical computers. Among its many applications, quantum sensing stands out for its ability to achieve unprecedent precision in measuring physical quantities. A critical aspect of quantum sensing is the optimization of entanglement distribution within quantum circuits, as entanglement is a fundamental resource that enhances the sensitivity and accuracy of quantum measurements [2]. Quantum entanglement is a unique feature of quantum mechanics that allows particles to exhibit correlations that are not possible in classical systems. In the context of quantum sensing, entanglement is leveraged to improve the precision of measurements, making it a valuable resource for quantum technologies. Quantum Fisher Information (QFI) is a measure of the sensitivity of a quantum state to changes in a parameter, and it plays a crucial role in assessing the performance of quantum sensors. Quantum sensing has emerged as a promising field that leverages quantum mechanical properties to achieve unprecedented levels of sensitivity and precision in measurement applications. One of the key challenges in quantum sensing is optimizing the distribution of entanglement within quantum circuits to enhance sensor performance while mitigating the effects of noise and decoherence. Quantum sensors exploit fundamental quantum phenomena such as superposition and entanglement to surpass the limitations of classical sensing technologies [1]. The distribution and manipulation of entanglement within a quantum circuit is therefore a key factor in optimizing sensor performance.

However, the practical implementation of quantum sensors faces several challenges, including the need to design and optimize complex quantum circuits that can maintain coherence and maximize entanglement in the presence of noise and environmental interactions[11].

Traditional approaches to quantum circuit optimization often rely on heuristic methods or manual design by experts. These methods can be time-consuming and may not fully exploit the potential of quantum resources [4]. Optimizing quantum circuits involves finding the best arrangement and sequence of quantum gates to achieve a desired outcome while minimizing resources and error accumulation. This is a complex task due to the vast search space of possible circuit configurations and the need to balance multiple, often competing, objectives such as maximizing entanglement, minimizing circuit depth, and maintaining robustness against noise.

This paper presents a novel an adaptive, automated approach to quantum circuit optimization using reinforcement learning techniques, specifically focusing on the optimization of entanglement distribution in quantum sensor circuits using sophisticated optimization techniques like entanglement injection, boosting, adaptive learning, entanglement layers.

## II. RESEARCH OBJECTIVE

This Research aims to develop a robust framework that uses QML techniques to dynamically optimize the distribution of entanglement layout in a Quantum sensor circuit to maximize the sensitivity and accuracy. Quantum Sensor network that uses the developed QML Algorithms to layout, distribute and maintain entanglement efficiently.

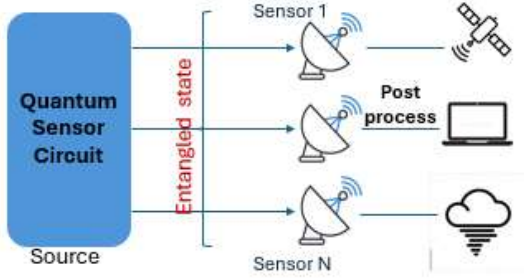

Fig. 1. Quantum sensor network interacting with external environment.

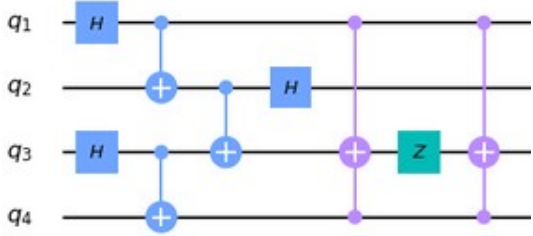

Fig. 2. Entanglement distribution in a quantum circuit.

## III. RELATED WORKS

*A. Greedy Algorithm Based Circuit Optimization for Near-Term Quantum Simulation.* In this work, they develop a circuit optimization algorithm to reduce the overall circuit cost, The method employs a novel sub-circuit two qubit synthesis in intermediate representation and proposes a greedy ordering scheme for gate cancellation to optimize the gate count and circuit depth. Gate Count Reduction by 7.8% and Depth Reduction by 16.3 %. The major approach is specific to the Hamiltonian circuits and complexity of greedy optimization is high and time consuming as the approach is greedy in nature.[5]

*B. Quantum Circuit Optimization with Deep Reinforcement Learning.* This is an approach to quantum circuit optimization based on RL. It demonstrates the feasibility of approach by training agents on 12 qubits random entangled circuit where on an average depth reduction by 27% and CNOT gate counts by 15%. Although the approach concentrates on 2 qubits gate optimization, the amount of entanglement is not measured during the process.[6]

*C. Reinforcement Learning based Quantum circuit optimization via ZX-Calculus.* It proposes a method for optimizing quantum circuits using graph-theoretic simplification rules of ZX-diagrams.
It illustrates its versatility by targeting both total and two-qubit gate count reduction, conveying the potential of tailoring its reward function to the specific characteristics. Here the work is better compared to other ZX calculus and heuristic algorithms. It does not focus on the sensitivity while optimizing the circuits .[7]

*D. Cost Explosion for Efficient Reinforcement Learning Optimization of Quantum Circuits.* The goal is to improve the agent's optimization strategy, by including hints about how quantum circuits are optimized manually: there are situations when the cost of a circuit should be allowed to temporary explode, before applying optimizations which significantly reduce the circuit's cost. The reward is a function of the quantum circuit costs, such as gate and qubit count, or circuit depth, entropy.[8]

*E. Size Optimization of CNOT Circuits on NISQ.* They study the optimization of the CNOT circuits on some noisy intermediate-scale quantum (NISQ) devices. Work is decomposed into two sub-problems: optimization with a given initial qubit distribution and optimization without limitations of initial qubit distribution. Though it focuses on two qubit gates, entropy is not measured.[9]

## IV. METHODOLOGY

The methodology employed in this entanglement optimization of quantum sensor circuit approach leverages reinforcement learning, specifically a Double Deep Q-Network (DDQN) agent, to optimize quantum circuits. The process begins with an initial quantum circuit, which is then iteratively modified by the DDQN agent. The agent learns from a deep Convolution network where it chooses between several circuit transformations or to generate another logically equivalent circuit. The process is repeated to achieve the best Reward.

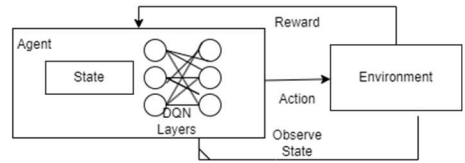

Fig. 3. Reinforcement learning framework with DDQN Network layers

The quantum state is represented using the state vector formalism. For an n-qubit system, the state is a complex vector in a $2^n$ dimensional Hilbert space.

The environment is modeled as a Markov Decision Process (MDP) where the state is represented by the current quantum circuit configuration. The circuit state is encoded as a 2D matrix of shape(M,5N+3), where M is the maximum number of gates and N is the number of qubits. Each row encodes one gate using one-hot vectors for gate type (H, CX, RX, RZ, CZ, SWAP, CRX) and gate parameters. The last 3 features include average layer entanglement, current entanglement, normalized depth, and normalized gate count. Thus, the RL agent receives a hybrid feature vector combining gate-level encoding and entanglement/depth metrics.

The DDQN algorithm is used to learn the optimal policy for circuit modification. Basically, it employs two neural networks: the main network for action selection and modification and the target network for stable Q-value estimation which helps to mitigate the over-estimation problems of agent actions. The Q-value update follows the equation:

$$Q(s,a) \leftarrow Q(s,a) + \alpha[r + \gamma Q'(s', argmax\_a\ Q(s',a)) - Q(s,a)] \quad (1)$$

where Q and Q' are the main and target networks respectively, α is the learning rate, γ is the discount factor, r is the reward, and s and s' are the current and next states. In addition to these layers, we also added experience reply to break correlations in the observation sequence and smooth over changes in the data distribution, experiences (s, a, r, s') are stored in a replay buffer and sampled randomly for training. A separate convolution layer to focus on entanglement and critic network layer to evaluate the value of state-action pair to help the agent to differentiate the action of benefits during the learning process. The environmental actions correspond to various circuit modifications such as adding gates, removing gates, entanglement injections, entanglement boosting and swapping gate positions. The RL agent learns to navigate this action space with a set of gates like Hadamard to create superposition that internally allows the qubit to exist in multiple states simultaneously to improvise the sensitivity of quantum sensor circuits. Entanglement gates such as CNOT, CZ, SWAP gates were used to create the entangled state along with few rotational gates like RX and RZ with parameter angle theta to rotate around x-axis and z-axis of the bloch sphere respectively, to maximize the precision of the entangled quantum sensor circuit.

### A. OPTIMIZATION TECHNIQUES

The agent uses an *epsilon-greedy strategy* for exploration, where it chooses a random action with probability ε and the best-known action with probability 1-ε. The value of ε is annealed over time to transition from exploration to exploitation. An *adaptive learning rate*

---

**Algorithm 1:** Entanglement Distribution optimization for QSC
**Input:** Initial quantum circuit C, hyperparameters: learning rate α, discount factor γ, exploration rate ε
**Output:** Entanglement optimized quantum circuit C'
1: **Initialization:**
   1.1: Initialize gate set G = {H, RX, RZ, CNOT, CZ, SWAP}
   1.2: Define the DDQN agent with a neural network for
       Approximation of Q-value function
   1.3: Set hyperparameters α, γ, ε
2: **Training Loop:**
  2.1: **for each** episode do
  2.2: Reset: Generate a random initial quantum circuit C or take the
      Input from user
  2.3: Convert C to a state vector s
  2.4: **for each** step within the episode do
  2.5: Select an action a using ε-greedy policy:
      2.5.1: With probability ε, select a random action
      2.5.2: Otherwise, select a = argmax_a Q(s, a; θ)
  2.6: Apply the action to C to obtain the next circuit C'
  2.7: Optimize the circuit C'
  2.8: Convert C' to the next state vector s'
  2.9: Calculate the reward R based on the Multi Reward function of C'
  2.10: Store the transition (s, a, r, s') in the replay buffer
  2.11: **if** the replay buffer is sufficiently full then
      2.11.1: Sample a minibatch from the replay buffer
      2.11.2: Update the Q-value network by minimizing the loss:
      2.11.3: L = (r + γ * max_{a'} Q(s', a'; θ-) - Q(s, a; θ))^2
      2.11.4: Update the target network θ- periodically
  2.12:    end if
  2.13:    s ← s'
  2.14:  end for
  2.15: end for

---

*scheduler* is implemented to adjust the learning rate based on the agent's performance, helping to stabilize and potentially speed up training. Adaptive learning uses *Adam optimization* technique to dynamically adjust the learning process based on the current performance or state of the system.

An *attention mechanism* allows the model to focus on different parts of the input state when making decisions. In the context of quantum entanglement optimization, it focuses on specific entangled gates and qubit interactions within the circuit.

$$\text{Attention}(Q, K, V) = \text{SoftMax}(QK^T/\sqrt{d_k})V \quad (2)$$

where Q, K, V, dk are query, key, value metrices derived from the input state and dimension of the keys.

*Entanglement Focused Layers* in the neural network architecture are designed to pay special attention to the entanglement properties of the quantum circuit. They might use custom activation functions or layer structures that are particularly sensitive to changes in entanglement. The system employs an *adaptive entanglement threshold* that adjusts based on the current circuit's entanglement level. This ensures that the optimization process maintains a minimum level of entanglement while allowing for circuit simplification.

*Periodic Entanglement Boosting*, if the circuit's entanglement falls below the threshold, the system injects additional entanglement by adding Hadamard and entangled gates. This technique helps maintain quantum advantage throughout the optimization process. Entanglement Boosting involves periodic reinforcement of entanglement across the circuit when the overall entanglement entropy drops below the threshold (set to 0.7). Boosting is performed by adding additional gates in specific regions to maintain high sensitivity.

*Layer-wise Entanglement Analysis* includes functionality to analyze entanglement at each layer of the circuit. This allows for targeted entanglement injections in layers where it's most needed by analyzing the weakly entangled layers, potentially improving the overall circuit performance.

*A circuit simplification* step is included after each action to reduce circuit complexity without significantly impacting performance. This likely involves techniques such as gate cancellation, replacements, sequencing and commutation rules. The system *dynamically updates error rates* for single-qubit and two-qubit gates based on the actions taken. This adaptive approach allows the optimization process to account for accumulated errors and noise effects more accurately.

*Entanglement injection* is a crucial technique used in this quantum circuit optimization approach to maintain or increase the level of quantum entanglement in the circuit. We identify injections as a specific gate pattern consisting of a Hadamard (H) followed by a CNOT, dynamically inserted into layers identified as weakly entangled during training. The adaptive and targeted nature of these injections allows the system to balance between circuit simplification and maintaining quantum resources effectively.

*Convergence-based Stopping in Optimization* by using a patience period, the algorithm doesn't stop immediately if progress slows down temporarily. The convergence check ensures that the optimization stops when both the entanglement and the circuit structure have stabilized.

This comprehensive approach combines techniques from quantum information theory, reinforcement learning, and circuit optimization to create an adaptive system for quantum circuit design and optimization. These advanced techniques work together to create a sophisticated reinforcement learning system capable of navigating the complex optimization landscape of quantum circuits. The critic network evaluates actions, the attention mechanism focuses on important circuit elements, adaptive learning adjusts the learning process, entanglement injection and boosting focuses on weakly entangled areas and inject the entangling gates to boost the correlation, entanglement focused layers capture quantum-specific features, and DDQN helps in more accurate Q-value estimation. All these components contribute to a robust and efficient quantum circuit optimization process.

### B. MULTI REWARD FUNCTIONS

A reward function in reinforcement learning is a crucial component that guides an agent's behavior by assigning a numerical value to its actions within a given state. This value, representing a reward or penalty, influences the agent's decision-making process as it strives to maximize its cumulative reward over time. As in our scenario, single reward function is not sufficient to meet the realistic challenges to achieve optimization with high precision. To address these challenges, we have incorporated multi reward function with different weights to balance the agent goal while optimizing the entanglement and circuit complexity reduction. The reward function is a weighted sum of improvements in QFI, depth reduction, entanglement enhancement, and gate count reduction:

$$R = w_1 * \Delta QFI + w_2 * \Delta Depth + w_3 * \Delta Entropy + w_4 * \Delta Gates \quad (3)$$

where $\Delta$ represents the change in each metric, and $w_1, w_2, w_3, w_4$ are weight parameters with 50,30,10,10 as weights considered after careful evaluation to prioritize the entanglement.

*1. Quantum Fisher Information (QFI):* It is used as a key metric for optimization by measuring the Sensitivity of the Quantum state for the whole Circuit over small perturbation of angle theta.

$$Q(\theta) = 4 * [\langle \partial_\theta \psi(\theta) | \partial_\theta \psi(\theta) \rangle - |\langle \partial_\theta \psi(\theta) | \psi(\theta) \rangle|^2] \quad (4)$$

where $|\partial_\theta \psi\rangle$ is the derivative of the state $\psi$ with respect to $\theta$. Higher the value of $Q(\theta)$, higher is the sensitivity. In this work, parameterized rotations $RX(\theta), RZ(\theta)$ are applied with the sensing parameter fixed at $\theta = \pi/2$ to ensure consistency with quantum sensing protocols.

*2. Entanglement Entropy:* The entanglement is quantified using the von Neumann entropy of the reduced density matrix. For a bipartite system AB, the entanglement entropy is:

$$S(\rho A) = -Tr(\rho A \log_2 \rho A) \quad (5)$$

where $\rho A$ is the reduced density matrix of subsystem.

*3. Depth Ratio:* Depth ratio is used as one of the rewards to measure the Depth Change and focuses on minimization of Circuit Depth[12]. This reward helps in reducing the complexity of overall circuits by keeping the state of entanglement unchanged.

$$\text{Depth Ratio} = D_{in} - D_{out} / D_{in} \quad (6)$$

where $D_{in}$ is the depth of input and $D_{out}$ is the depth of output circuit.

*4. Gate Ratio:* Gate ratio is used as one of the rewards to measure the gate counts change and focuses on minimization of gate counts. This reward helps to reduce the noise of overall circuits by cancelling noisy and unwanted gates.

$$\text{Gate Ratio} = G_{in} - G_{out} / G_{in} \quad (7)$$

where $G_{in}$ is the gates of input and $G_{out}$ is the gates of output circuit.

## V. EXPERIMENTS AND RESULTS

The experiment was designed to evaluate the effectiveness of the Deep Reinforcement Learning (DRL) approach in optimizing entanglement of quantum circuits. The core of the experiment involves training a Double Deep Q-Network (DDQN) agent to modify quantum circuits with the goal of improving entanglement distribution and other key metrics while maintaining or enhancing circuit functionality. (DDQN) agents demonstrate superior performance in quantum circuit optimization compared to alternative reinforcement learning approaches like overestimation, dual network structure to handle actions, Q-value evaluation, and ability to mitigate noise. A custom Quantum Circuit Environment was created using OpenAI Gym, simulating a quantum circuit with a specified number of qubits from 2 to 20 and a maximum number of gates from 15 to 160 respectively. The agent was trained for 1000 episodes, with each episode starting from one of the loaded initial circuits. To provide more flexible action space we have added many actions like the agent could perform various actions including adding different types of quantum sensing gates (H, CNOT, RX, RZ, CZ, SWAP), removing gates, swapping gate positions, gate cancellation, injection, boosting and replacing gates. Also, by keeping DDQN as the base network layer we have experimented with two additional network layers, one is the Critic network layer to evaluate the agent actions and entanglement focused network layer to understand the behavior of entangled gates by taking previous entanglement features as an input. The setup of the environment and hyperparameters used during the training are given in detail below.

| Elements | Value | Description |
|---|---|---|
| Packages used | Qiskit, TensorFlow, keras, NumPy, gym, matplotlib, sklearn, pandas | All the classical and quantum packages that are used to process the proposed DDQN approach |
| Training episodes | 1000 | The size of the training episodes to train the agent |
| Qubits | 2,3,5,8,10,15,20 | The size of the testing dataset to test the HCQNN model |
| Quantum simulator | AER, state vector | Simulators used for simulation |
| Memory size Batch size | 2000 64 | Maximum number of experiences sampled from the reply buffer |
| Discount factor(gamma) | 0.95 | Used in Q-value update |
| Epsilon Decay | 0.999 | The rate at which the exploration rate (ε) is decreased |
| Entanglement threshold | 0.7 | The minimum entanglement level required for the circuit. |

TABLE 1. The setup and hyperparameters of the environment

Initially we started the experiment by training the agent with 5 qubit circuit to analyze the state of DDQN architecture. As the number of gates and qubits are minimal for computation the results are bit less due to minimal opportunities for exploration. Moving forward we trained the agent for 8,10,15 and 20 qubits with higher number of gates upto 160 to provide more action space for exploration and exploitation. The results were quite satisfactory and we are able to achieve the goal of balancing the multi reward functions by maximizing the QFI and entanglement of the circuit by minimizing the complexity of circuit with depth and gate counts reduction. Before applying different optimization techniques we were able to optimize the entanglement layout with maximum QFI and entropy of average 0.80 to 0.90 with average depth reduction of 25 to 78 % and gate counts reduction of 23 to 81% on an average. But as our main goal is to achieve the high entangled circuit we have adopted few optimization techniques to focus on the area of entanglement within the quantum sensor circuit. Then after applying the previously mentioned entanglement concentrated optimization techniques the results have been improved with maximum QFI and entropy of 0.84 to 1.0 with average depth and gates count reduction of 20-81%. In order to set the benchmark and compare our work with other related works we have choosen one of the closest work [7] as the work emphasize in optimizing entangling gates with RL approach and is readily avaiable to consider it for comparision as a baseline model [14].

we have modified the base implementation by adding few measuring metrics to make it compaitable for comparison with our work.As a result we are able to achieve high entanglement by minimizing depth and gate counts by 4 times more than the baseline model.

To support the realistic environment we have injected a realistic noise model using qiskit's noise model class and simulated using Aersimulator under noisy environment.The noise model includes depolarizing errors for measurement and gates and thermal relaxation errrors modeling T1 and T2 processes.

The inclusion of this noise model allows the optimization process to generate circuits that are more robust to real-world quantum hardware limitations.As per the expectations the results are quite compromising in the noisy environment while comparing with noiseless model.The quantum circuit entanglement optimization algorithm demonstrates exponential time complexity $O(2^n)$, where n represents the number of qubits, reflecting the inherent computational challenges in quantum circuit optimization. When implemented on a local machine with 8GB RAM and 1TB HDD specifications, the algorithm requires approximately $2^n$ hours to converge to optimal solutions, highlighting the significant computational resources demanded by quantum optimization tasks.

These experiments and results demonstrate the potential ability of the DRL approach to optimize the entanglement distribution for qunatum sensor circuits by showing improvements across multiple objectives simultaneously.While adapting to increasing noise levels highlights the method's strength for practical algorithms design and optimization. Pareto plots and reward curves serve as essential analytical tools in tracking,helps to balance the trade-off between multiple reward functions and monitor the learning process of agent within the boundary of optimization process that determines the strength and stability of the agent both in noisy and noiseless environments. The visualizations and tables related to the works are as follows below.

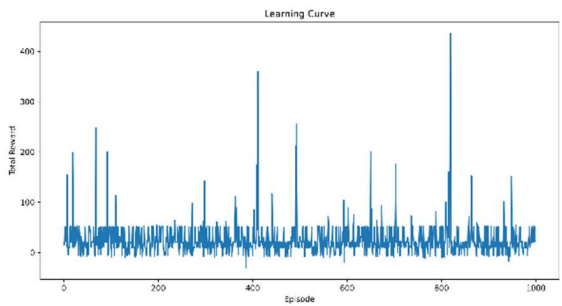

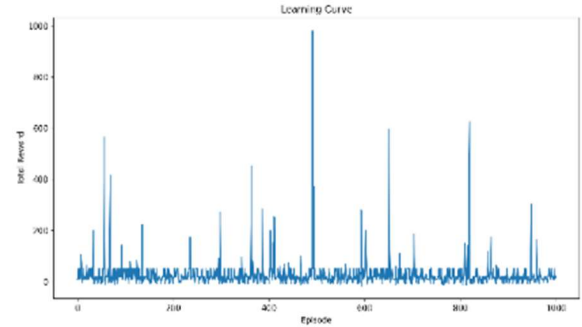

Fig. 4. Total reward curve of 10 qubit circuits with multi metrics under noiseless environment over episodes shows overall rewards achieved by the agent while optimizing the circuit to reach maximum entanglement.

Fig. 5. Total reward curve of 10 qubit circuits with multi metrics under noisy environment over episodes shows overall rewards achieved by the agent while optimizing the circuit to reach maximum entanglement under noisy environment which defines the stability of model when compared with fig 4.

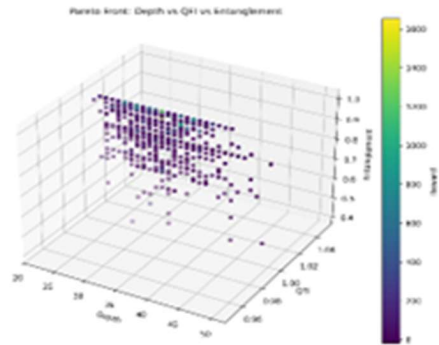

Fig. 6. Pareto plot of 10 qubit circuits with multi metrics under noiseless environment.

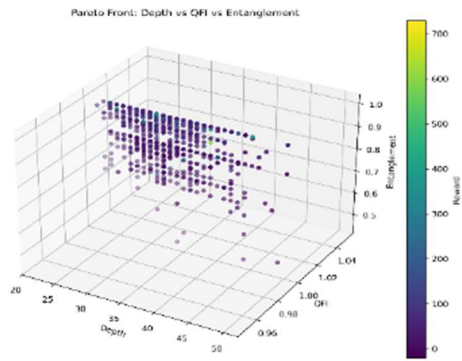

Fig. 7. Pareto plot of 10 qubit circuits with multi metrics under noisy environment.

Pareto plot over episode,helps to monitor the balanced tradeoff between the multi rewards in achieving high QFI and Entanglement while minimizing the depth and gate counts of the quantum circuit.

The line graphs below display the results of an experiment involving entanglement optimization across 1,000 different quantum circuits. The x-axis represents the number of circuits tested, while the y-axis represents the optimization rewards, specifically Quantum Fisher Information (QFI) and Entropy. The observed spikes, which rise from a lower point to a peak, indicate successful maximization of entanglement within the corresponding circuits.

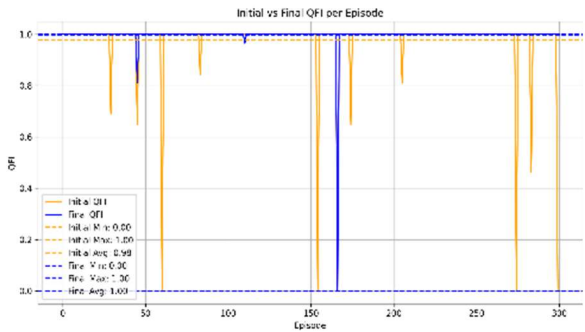
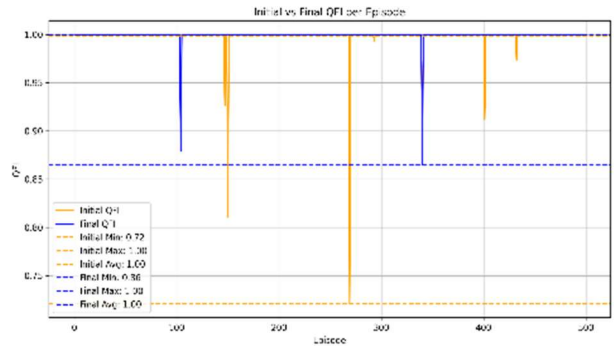

8a                                                                    8b

Fig. 8a. The visualization of Quantum Fischer (QFI) of 8 Qubit from 0.98 to 1.0    Fig. 8b. The visualization of Quantum Fischer (QFI) of 10 Qubit circuit with highest value of 1.0

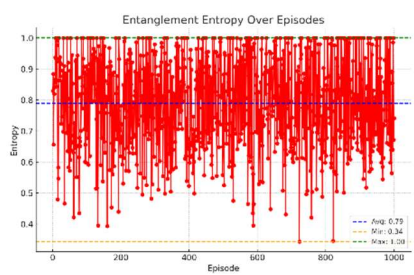
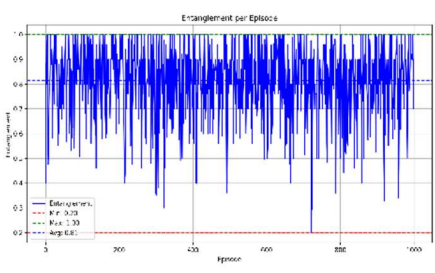
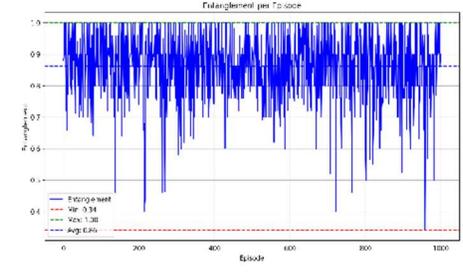

9a                               9b                                9c

Fig 9. The visualization of Entanglement entropy of 10 Qubit circuit. 9a) The Entropy of zx-calculus 9b) The Entropy of DDQN model before applying optimization techniques 9c) The QFI of DDQN model after applying optimization techniques

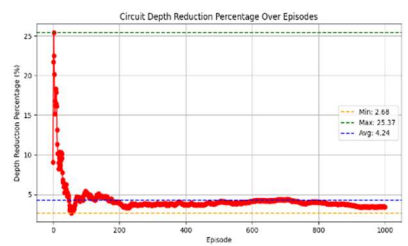
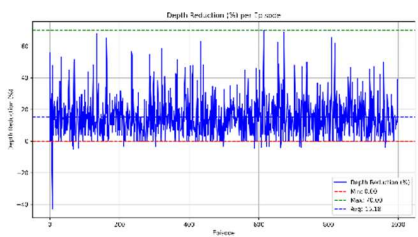
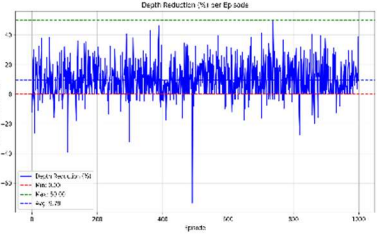

10a                              10b                               10c

Fig 10. The visualization of Depth Reduction of 10 Qubit circuit. 10a) The depth reduction of zx-calculus 10b) The depth reduction of DDQN model before applying optimization techniques 10c) The depth reduction of DDQN model after applying optimization techniques

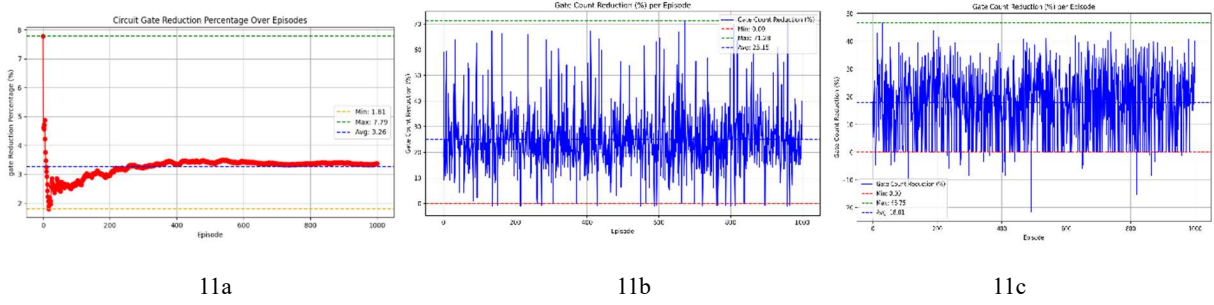

11a            11b            11c

Fig 11. The visualization of Gate Reduction of 10 Qubit circuit. 11a) The gate reduction of zx-calculus 11b) The gate reduction of DDQN model before applying optimization techniques 11c) The gate reduction of DDQN model after applying optimization techniques

| Qubit | Max gates | QFI | | Entropy | | | | Depth Reduction % | | | | Gates Reduction % | | | |
|---|---|---|---|---|---|---|---|---|---|---|---|---|---|---|---|
| | | Base line | final | Base line | i | ii | iii | Base line | i | ii | iii | Base line | i | ii | iii |
| 5 | 60 | 1.0 | 1.0 | 0.86 | 0.85 | 0.89 | 0.88 | 16.48 | 12.17 | 5.08 | 4.97 | 13.94 | 20.03 | 11.47 | 11.55 |
| 8 | 90 | 1.0 | 1.0 | 0.92 | 0.90 | 0.92 | 0.91 | 8.29 | 12.35 | 10.04 | 9.7 | 4.68 | 17.72 | 17 | 16.84 |
| 10 | 90 | 1.0 | 1.0 | 0.79 | 0.81 | 0.86 | 0.85 | 8.48 | 15.18 | 9.79 | 9.84 | 6.52 | 25.15 | 18.01 | 20.60 |
| 15 | 120 | 1.0 | 1.0 | 0.85 | 0.80 | 0.81 | 0.81 | -2.33 | 9.06 | 10.05 | 9.84 | 2.11 | 17.93 | 18.69 | 18.42 |
| 20 | 160 | 1.0 | 1.0 | 0.85 | 0.80 | 0.82 | 0.76 | -3.89 | 13.16 | 7.15 | 6.85 | 1.72 | 13.26 | 13.37 | 13.26 |

TABLE 2. The Results table of 5 to 20 Qubits Quantum circuit using Qiskit compiler with multi metrics QFI, Entropy, Depth Reduction and Gate counts Reduction compared to the baseline work [7] i) Result values estimated before applying optimization techniques. ii) Result Values estimated after applying optimization techniques. iii) Result values estimated with optimization techniques under the noisy environment. The Results clearly state that the model can achieve highest QFI and Entanglement Entropy of average 0.80 to 0.92 and max of 1.0 along with the average depth and gate counts reduction by 25% and maximum by 86%.

| Related Work | Measures | | | | | | | |
|---|---|---|---|---|---|---|---|---|
| | Machine Learning | QFI | Entropy | Depth Reduction | Gate Counts Minimization | Entanglement Boosting | Entanglement Injection | Flexibility |
| Greedy Algorithm Based Circuit Optimization [5] | X | X | X | ✓ | ✓ | X | X | X |
| Quantum Circuit Optimization with Deep Reinforcement Learning. [6] | ✓ | X | X | ✓ | ✓ | X | X | X |
| Reinforcement Learning Based Quantum Circuit Optimization via ZX-Calculus.[7] | ✓ | X | X | ✓ | ✓ | ✓ | X | ✓ |
| Cost Explosion for Efficient Reinforcement Learning Optimization of Quantum Circuits [8] | ✓ | X | ✓ | ✓ | ✓ | ✓ | X | X |

| Size Optimization of CNOT Circuits on NISQ(V3) [9] | X | X | X | ✓ | ✓ | X | X | ✓ |
| Quantum Machine Learning for Optimizing Entanglement Distribution in Quantum Sensor Circuit | ✓ | ✓ | ✓ | ✓ | ✓ | ✓ | ✓ | ✓ |

TABLE 3. The table represents the qualitative analysis of our work in comparison with other related works by considering few key measuring criteria to highlight the efficiency and effectiveness of improvisation in our approach.

We have also experimented with Tket compiler to automate the process of Depth and gate counts optimization. To optimize the gate and depth counts by preserving the state of sensitivity we merged the qiskit passes and Tket passes along with our custom templates to choose the best optimized circuit by comparing the measurement values. The custom Tket sequence passes try to simplify the complexity of circuit by converting all single qubits circuits into equivalent parameterized universal gate by commuting the entangled gates into the right position.

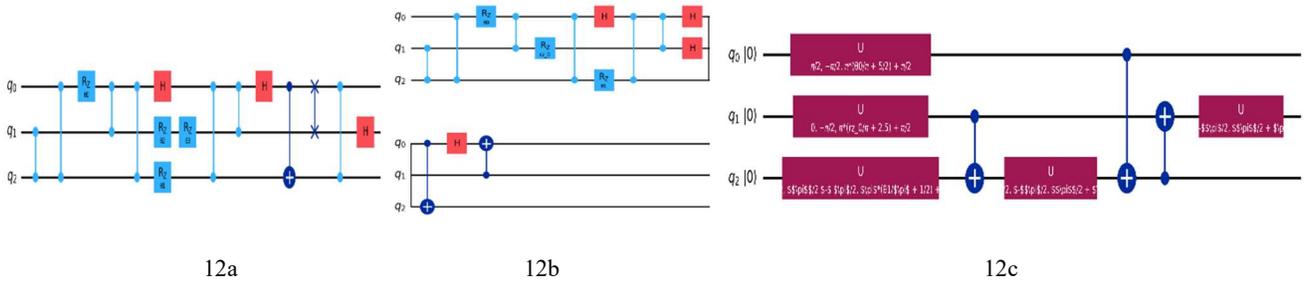

12a      12b      12c

Fig 12. 12a) Randomly generated input circuit with depth 12, 0 QFI and Entropy 12b) Qiskit optimized circuit with no change in sensitivity 12c) Tket optimized circuit with depth 8 and maximum QFI and entanglement entropy of 1.0 along with gate counts reduction to 8.

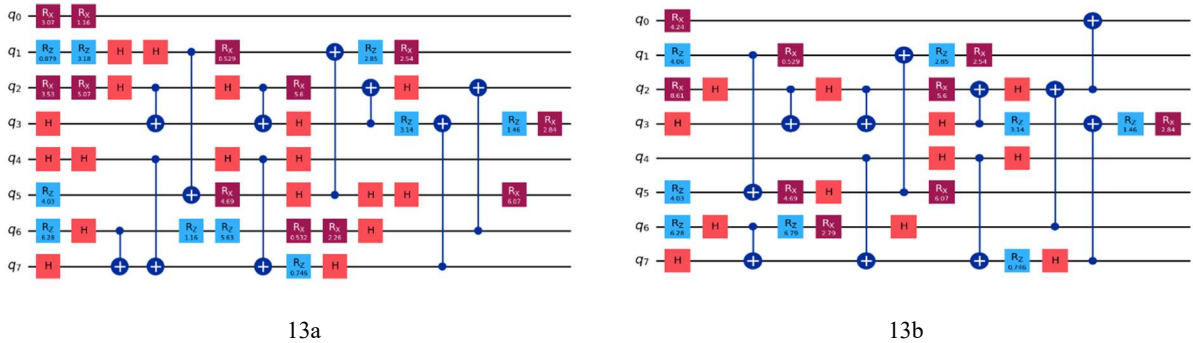

13a      13b

Fig 13. 13a) Input circuit with entropy 0.4759, depth of 12 with 50 gates. 13b) Qiskit Optimized circuit with entropy 0.6836, depth of 11 with 40 gate counts. Entanglement is improved by 0.2077 with gate change of 10 counts.

| Qubits | Max gates | QFI | | Entropy | | Depth Reduction % | | Gates Reduction % | |
|---|---|---|---|---|---|---|---|---|---|
| | | Initial | Final | Initial | Final | Max | Avg | Max | Avg |
| 2 | 15 | **0.89** | **0.93** | 0.43 | 0.83 | 85.71 | 42.13 | 85.71 | 36.82 |
| 3 | 30 | **0.93** | **0.97** | 0.55 | 0.82 | 85.71 | 30.48 | 80.65 | 29.17 |
| 5 | 60 | **0.72** | **0.80** | 0.81 | 0.99 | 62.50 | 7.80 | 60 | 12.19 |
| 8 | 90 | **0.98** | **1.0** | 0.64 | 0.80 | 62.50 | 9.27 | 60 | 19.75 |
| 10 | 120 | **1.0** | **1.0** | 0.76 | 0.86 | 57.14 | 7.57 | 56.67 | 19.63 |

TABLE 4. The Results table of 2 to 10 Qubits Quantum circuit with multi metrics QFI, Entropy, Depth Reduction and Gate counts Reduction. Using both qiskit and Tket compiler together for the automation of depth and gate counts optimization. Initial value represents the measurements of input circuits sent for entanglement distribution optimization. Final Value column represents the measurements of optimized circuits in terms of different measuring metrics. The result states that the model can optimize the entanglement distribution for quantum sensor circuits by demonstrating the maximum QFI and Entanglement entropy of 0.99 by reducing the complexity of circuits with an average of 42% and maximum of 85.71% in depth and gate counts reduction.

## VI. Conclusion and future work

The research effectively adopted Deep Reinforcement Learning techniques to dynamically optimize the entanglement distribution of quantum sensor circuits. The approach provides a robust framework for enhancing quantum circuit sensitivity and performance by optimizing key metrics such as entanglement, depth, and gate count by adopting unique optimization techniques like entanglement injection, boosting, adaptive learning, layer wise optimization etc.

The results demonstrate the model's ability to achieve high sensitivity with maximum Quantum Fisher Information (QFI) and entanglement entropy in the range of 0.8 to 1.0, along with a 20-86% reduction in circuit depth and gate count. This efficiency in managing high entanglement with minimum depth and gates is particularly impressive for 2-20 qubit circuits.

The framework and protocols developed in this work can be scaled to larger quantum sensor networks, paving the way for practical implementations of quantum sensor networks in various applications like metrology and others. The resulting circuits exhibit enhanced sensitivity, reduced complexity, and improved robustness to noise – all of which are crucial for the successful deployment of quantum technologies in real-world sensing and metrology applications. The insights and methodologies presented in this work can serve as a valuable foundation for future developments in the field of quantum computing and quantum sensor design.

Our future work will focus on improving the scalability of the code to handle larger quantum circuits with more qubits and gates. Focus on more appropriate error mitigation mechanisms to mitigate the effects of noise and errors in quantum circuits. To experiment with various neural network architectures like PPO Agents and Policy learning. Extending our work to test and validate the optimized quantum circuits on real quantum hardware, exploring performance metrics under practical conditions. The optimization algorithm exhibits exponential complexity with respect to the number of qubits. While simulations up to 20 qubits were successful, scaling to larger systems remain challenging. To overcome this scalability issue, we have planned to integrate tensor network techniques, particularly the Matrix Product State (MPS) approach, which significantly reduces simulation complexity for larger circuits. Additionally, future work will explore hardware-aware optimizations to further improve scalability.

## Acknowledgments

This work was supported by the National Science Foundation Grant No. OMA 2231377.

Provincial Key Laboratory of Quantum Metrology and Sensing & School of Physics and Astronomy.